\documentclass[a4paper,twocolumn,preprintnumbers,superscriptaddress,citeautoscript,newabstract,aps,nofootinbib]{revtex4-2} 
\usepackage{amsmath,amsfonts,amssymb,graphics}
\usepackage{dsfont}
\usepackage[normalem]{ulem}
\usepackage{graphicx}
\usepackage{units}
\usepackage{mathrsfs}
\usepackage{bbm}
\usepackage{pifont}
\usepackage{braket}
\usepackage{units}
\usepackage[dvipsnames]{xcolor}
\usepackage{mathtools}
\usepackage{xspace}
\usepackage{booktabs}
\usepackage{slashed}
\usepackage{feynmp-auto}
\usepackage[bookmarks=false,hyperfootnotes=false]{hyperref}

\newcommand{\SO}[1]{\ensuremath{\mathrm{SO}(#1)}}

\newcommand{\U}[1]{\ensuremath{\mathrm{U}(#1)}}

\renewcommand{\bar}[1]{\overline{#1}}

\begin{document}
\title{The goofy-symmetric Standard Model and the Hierarchy Problem}

\author{Thede de Boer}\email[]{thede.deboer@mpi-hd.mpg.de}
\affiliation{\vspace{0.2cm}Max-Planck-Institut f\"ur Kernphysik, Saupfercheckweg 1, 69117 Heidelberg, Germany}
\author{Florian Goertz}\email[]{florian.goertz@mpi-hd.mpg.de}
\affiliation{\vspace{0.2cm}Max-Planck-Institut f\"ur Kernphysik, Saupfercheckweg 1, 69117 Heidelberg, Germany}
\author{Andrea Incrocci}\email[]{andrea.incrocci@mpi-hd.mpg.de}
\affiliation{\vspace{0.2cm}Max-Planck-Institut f\"ur Kernphysik, Saupfercheckweg 1, 69117 Heidelberg, Germany}

\begin{abstract}
A new class of so-called ``goofy'' symmetries has been shown to lead to renormalization-group stable relations between parameters in two-Higgs-Doublet Models, not known before. 
In this work we investigate goofy transformations in the Standard Model (SM), extending them to the fermion sector.
We show that the SM action is goofy ``symmetric'', if the Higgs mass term vanishes, while other SM interactions are allowed. Goofy symmetry could thus add another option to the very few means to successfully forbid the Higgs mass term, providing an alternative to conformal symmetry or shift symmetry.
We further consider how electroweak symmetry breaking can be realized in the goofy symmetric SM via higher dimensional operators or an extended scalar sector. A viable Higgs mass can be obtained via spontaneous breaking of goofy symmetry, which can be realized via operators that would not be allowed by shift symmetry or conformal symmetry. We find that goofy-symmetric extensions of the SM are strongly constrained, but offer for example a straightforward dark matter candidate.

\end{abstract}

\maketitle

\section{Introduction}

Recently, new renormalization-group (RG) stable parameter relations in two-Higgs-Doublet Models (2HDMs), that do not correspond to any known symmetry, have been discovered~\cite{Ferreira:2023dke}.
They have accordingly been related to invariance of the scalar potential under novel so-called {\it goofy} transformations~\cite{Ferreira:2023dke} -- where the name reflects both their unusual nature and the initials of the authors~\cite{FerreiraTalk}.
Indeed, while imposing these parameter relations does not realize any traditional symmetry of the 2HDM, one can identify a corresponding peculiar transformation -- to be discussed further below -- that leaves the Lagrangian invariant up to a sign-flip in the scalar kinetic terms. Similar to traditional symmetries, these ``goofy symmetries'' lead to parameter relations that are stable under the RG-flow, as one can show from general arguments, various aspects of which have been put forward and explored in~\cite{Trautner:2025yxz}. There it has also been speculated that the fact that goofy symmetries can forbid bare mass terms could be relevant for tackling the hierarchy problem.
Moreover, in Ref.~\cite{Haber:2025cbb}, it has been analyzed how goofy symmetries could be understood by ``complexifying'' the theory. In the theory where each real scalar degree of freedom (dof) is augmented to a complex scalar, goofy symmetries correspond to regular symmetry transformations, guaranteeing corresponding parameter relations. The transformation of the 1-loop effective potential of the 2HDM under goofy transformations has been considered in Ref.~\cite{Ferreira:2025ate}, finding that the effective potential remains invariant, if the cutoff (or the RG scale) also transforms accordingly (see however \cite{Pilaftsis:2024uub}).

In this letter, we apply goofy symmetry to the Standard Model (SM) Higgs sector and extend it beyond the scalar potential to include the SM fermions. In particular, we demonstrate how in this context goofy symmetry could indeed explain the lightness of the Higgs boson. In fact, all SM interactions will be compatible with the extended goofy symmetry, while the Higgs-mass operator will be forbidden. We will show how still a non-trivial Higgs potential can be realized, generating a viable Higgs mass and vacuum expectation value (vev) via higher dimensional operators and consequently leading to spontaneous goofy symmetry breaking. Alternatively, goofy symmetry can be spontaneously broken in an extended scalar sector in which case the Higgs mass term is generated via the portal term.
Interestingly, the responsible operators would be forbidden in conformally symmetric models, but are not in conflict with goofy symmetry of the action. Finally, we will explore further possibilities for extensions of the SM that respect goofy symmetry.

\section{Goofy transformations in the SM}
Goofy symmetries can be understood as rotations of real scalar dofs in the complex plane by $90^{\circ}$, i.e.\ multiplying by a factor of $i$, while still restricting the fields to the resulting one-dimensional axis.
Writing the Higgs doublet in terms of its real components as
\begin{equation}
\label{eq:doubl}
    H=\left(\begin{matrix}
        h_1+ih_2\\h_3+ih_4
    \end{matrix}\right)\,,
\end{equation}
the most general goofy transformation can be written as 
\begin{equation}\label{eq:scalar trafo}
    h_i\to i\,U^{ij}_h h_j\,,
\end{equation}
for an \textit{orthogonal} matrix $U_h$. In the following, we consider $U_h={\bf 1}$
which, if grouping the dofs as in \eqref{eq:doubl}, is equivalent to the replacement
\begin{equation}
H\to i H,\quad H^\dagger\to i H^\dagger\,,
\end{equation}
in the Lagrangian.
In general, goofy transformations do not leave the kinetic term invariant, but transform it as
\begin{equation}
    \mathcal{L}_\mathrm{kin}=(D_\mu H)^\dagger(D^\mu H)\to-(D_\mu H)^\dagger(D^\mu H)\,.
\end{equation}
However, it has been speculated that goofy symmetries may be promoted to exact symmetries of the action, leaving also the kinetic terms invariant, if the derivative operator and gauge fields transform as
\begin{equation}
    \partial_\mu \to -i \,\partial_\mu\,,\quad A_\mu^a \to -i A_\mu^a\,, \label{eq. trafo Dmu}
\end{equation}
which would require transforming spacetime coordinates to imaginary values~\cite{Ferreira:2023dke}.\footnote{It would be interesting to further explore the possible interrelation of goofy symmetries with discrete space-time symmetries, such as PT, and their (potentially spontaneous) breaking, which is left for future work.} In fact, considering such transformation will turn out to be instrumental for a successful inclusion of fermions, see below.

The Higgs potential transforms as
\begin{equation}
    V=m_H^2 |H|^2+\lambda |H|^4\to-m_H^2 |H|^2+\lambda|H|^4\,,
\end{equation}
and therefore $m_H^2$ explicitly breaks the goofy symmetry. Small values are therefore natural in the 't Hooft sense and expected to be radiatively stable -- {\it if} the rest of the SM (and new physics) is respecting goofy symmetry.\footnote{Note that Higgs shift symmetry $H\to H + a$, which could also protect the Higgs mass, is broken considerably in the SM, for example by the Higgs self-coupling and the large top Yukawa interaction (see e.g.~\cite{Schmaltz:2005ky}).  For ideas of how to address this, see e.g. \cite{Bally:2022naz,Chung:2024zsr,Chung:2023gcm}.} We will explore this possibility in this and the next section.

The Yukawa interactions in the SM quark sector, with explicit hermitian conjugates and focusing on one generation, are given by
\begin{equation}
\begin{split}
\mathcal{L} \supset &  - y_d \bar Q_L  H d_R - y_u \bar Q_L \tilde H u_R \\ & - y_d^{*} \bar d_R H^\dagger Q_L- y_u^{*} \bar u_R \tilde H^{\dagger} Q_L\,.
\end{split}
\end{equation}
These operators are invariant under the goofy transformation, if the fermions transform as
\begin{equation}
\begin{split}
\bar Q_L&\to \sqrt i \, \bar Q_L\,,\quad \ \ \, Q_L\to \sqrt i\, Q_L\,,\\
d_R&\to -\sqrt i\, d_R\,,\quad \,\bar d_R\to -\sqrt i\, \bar d_R\,,\\
u_R&\to -\sqrt i\,u_R\,,\quad \, \bar u_R\to -\sqrt i\,\bar u_R\label{fermgoof}\,,
\end{split}
\end{equation}
which guarantees that the fermionic kinetic term transforms as
\begin{equation}
\begin{split}
\mathcal{L}_{\text{kin}}^{\text{ferm}} =& i \bar Q_L \slashed D Q_L + i \bar d_R \slashed D d_R + i \bar u_R \slashed D u_R\\
\to& i\left(i \bar Q_L \slashed D Q_L + i \bar d_R \slashed D d_R + i \bar u_R \slashed D u_R\right)=i\mathcal{L}_{\text{kin}}^{\text{ferm}}\,.
\end{split}
\end{equation}
This latter transformation is essential in order for the full Lagrangian to be invariant if combined with~\eqref{eq. trafo Dmu}. Analogous arguments apply for the lepton sector.

Like this, in line with the bosonic case, the transformations of the spinors can be understood as a rotation of the real degrees of freedom by $45^{\circ}$ in the complex plane: 
If we decompose a generic of spinor $\Psi$ in terms of real components $\psi^r_1$ and $\psi^r_2$ as
\begin{equation}
    \Psi=\psi^r_1+i\,\psi^r_2\,,
\end{equation}
then the goofy transformation acting on the real degrees of freedom of the fermions can be written as\footnote{We restrict ourselves to transformations that do not act on flavor and spinor indices.}
\begin{equation}\label{eq:trafo fermions}
    \psi^r_k\to \sqrt{i}\,U_f^{kl}\psi^r_l\,,
\end{equation}
for an \textit{orthogonal} matrix $U_f$ and with $k,l=1,2$. If, for example, we take $U_f$ as the standard rotation matrix in two dimension with an angle of rotation $\varphi$, this implies
\begin{equation}
\begin{split}
    \Psi\to\sqrt{i}\,e^{i\varphi}\Psi\,,\qquad\bar{\Psi}\to\bar{\Psi}\sqrt{i}\,e^{-i\varphi}\,,
\end{split}
\end{equation}
and the kinetic term then transforms as 
\begin{equation}
    \bar{\Psi}\slashed D\Psi\to i\,\bar{\Psi}\slashed D\Psi\,,
\end{equation}
and remains invariant if combined with~\eqref{eq. trafo Dmu}.
%
The transformations~\eqref{fermgoof} are retrieved with $U_f=\mathbf{1}$ for the doublet $Q_L$ and $U_f=-\mathbf{1}$ for the singlets $d_R$ and $u_R$.
It is exactly this consistent extension of goofy symmetry to the fermion sector that will guarantee the absence of corrections to the Higgs-mass term in the presence of new physics, see below.

While the Higgs mass term is prohibited by goofy symmetry, electroweak symmetry breaking (EWSB) can be realized, similar to Ref.~\cite{Goertz:2015dba}, by higher order operators such as\footnote{$|H|^6$ is forbidden by goofy invariance.}
\begin{equation}
\label{eq:D8}
V=\lambda |H|^4+\frac{c_8}{\Lambda^4}|H|^8\,.
\end{equation}
A realistic Higgs vev and Higgs mass can be obtained with
\begin{equation}
\lambda=-0.065\,,\quad c_8=4\pi\left(\frac{\Lambda}{773\,\mathrm{GeV}}\right)^4\,,
\end{equation}
which necessitates a UV completion of the $|H|^8$ operator below the TeV scale.
Further, this realization of EWSB yields modified Higgs self couplings. Explicitly, the triple Higgs coupling $\kappa_{hhh}$ will be given by $\kappa_{hhh}/\kappa^\mathrm{SM}_{hhh}=3$, while the four-Higgs coupling $\lambda_{hhhh}$ will become $\lambda_{hhhh}/\lambda^\mathrm{SM}_{hhhh}=17$. Both of these values are compatible with current experimental bounds \cite{ATLAS:2022jtk, ATLAS:2024xcs}, and could be tested at the HL-LHC and future colliders \cite{DiMicco:2019ngk}.
The one-loop running of the Higgs mass, calculated in Ref.~\cite{Helset:2022pde}, explicitly confirms that the operator $|H|^2$ will not be generated at loop level in the presence of the $\left|H\right|^8$ term, consistent with goofy-symmetry. It is interesting to note that this construction \eqref{eq:D8} preserves the goofy symmetry, but breaks conformal symmetry. Another option for realizing EWSB would be via the coupling to a scalar singlet, that we will discuss next, in Sec.~\ref{sec:singl}.

\section{Extensions of the SM}

\subsection{Scalar singlet}
\label{sec:singl}
Goofy symmetry might also be spontaneously broken in an extended scalar sector. In such a setting, the Higgs mass term would be generated by the portal term. We illustrate this by amending the SM with a complex scalar singlet $S=\frac{1}{\sqrt{2}}\left(S_R+iS_I\right)$. We choose the orthogonal matrix, analogous to $U_h$ in \eqref{eq:scalar trafo}, as $U_S=-i\sigma_2$, where $\sigma_2$ is the second Pauli matrix. 
This fixes the transformation of $S$ under goofy symmetries, similar to how one chooses representations for traditional symmetries.\footnote{The choice $U_S=\mathbf{1}$ does not allow any mass terms for $S$ and therefore does not lead to spontaneous symmetry breaking.} 
Then the scalar singlet $S$ transforms as 
\begin{equation}
S\to -S\,,\quad S^*\to S^*\,.
\label{scalar_goofy}
\end{equation}
If we impose the goofy symmetry (and hermiticity), the scalar potential is given by
\begin{equation}
\begin{split}
    V=&\frac{m_1^2}{2}\left[ S^2+ (S^*)^2\right]+\lambda_H |H|^4+\lambda_p |S|^2|H|^2\\
    &+\lambda_S |S|^4+\left[\lambda_1 S^4+ \lambda_1^* (S^*)^4\right]\,,
\end{split}
\label{eq:VS}
\end{equation}
where, without loss of generality, we assumed $m_1^2$ to be real.\footnote{As a sanity check, we confirmed that no new operators are generated along the RG flow using the two loop RGEs computed with \texttt{PyR@TE}~\cite{Sartore:2020gou}.} 
If we write $S$ in terms of its real components, $S=\frac1{\sqrt{2}}(S_R+iS_I)$, we find 
\begin{equation}
    V=\frac{m_1^2}{2}\left[ S_R^2- S_I^2\right]+\frac{\lambda_p}{2} (S_R^2+S_I^2)|H|^2+... \,.
    \label{eq:VSR}
\end{equation}
Assuming a small portal coupling and $m_1^2>0$, $S_I$ obtains a vev given by
\begin{equation}
    v_S^2=\frac{m_1^2}{2\mathrm{Re}(\lambda_1)+\lambda_S}\,,
\end{equation}
spontaneously breaking the goofy symmetry and generating  a Higgs mass term of $m_H^2=\lambda_pv_S^2/2$. For the case of $m_1^2<0$, the situation is inverted and $S_R$ assumes the vev (with $m_1^2\to-m_1^2$). We conclude that the scalar singlet masses are limited by the scale of spontaneous goofy symmetry breaking -- which, in the presence of a sizable Higgs portal, should not be too far above the electroweak scale in order not to reintroduce the hierarchy problem. 
Finally, it is interesting to note that if $\lambda_1$ is real, the singlet $S_R$ (for $m_1^2 >0$) is a Dark Matter candidate, being stable due to an unbroken $Z_2$ symmetry $S_R \to - S_R$ in~\eqref{eq:VS}.

\subsection{Vector-like top quark}

In order to explore how fermions can be added to the SM while respecting goofy symmetry, we consider a vector-like quark with components $T_L$ and $T_R$, being $SU(2)_L$ singlets with hypercharge $Y=\frac23$. Without imposing goofy symmetry, the Lagrangian is extended by
\begin{equation}
\begin{split}
\mathcal{L}\supset\ &m_{\cal T} \bar T_L T_R+m_{\cal T}^* \bar T_R  T_L+m \bar T_L u_R\\
    &+m^* \bar u_R T_L+ y \bar Q_L\tilde H T_R+y^*\bar T_R \tilde H^\dagger Q_L\,.
\label{eq:VL}
\end{split}
\end{equation}
The new Yukawa interaction is invariant, if $T_R$ transforms as (c.f. \eqref{fermgoof})
\begin{equation}
    T_R\to  -\sqrt{i}\, T_R\,,\qquad \, \bar T_R\to -\sqrt{i}\, \bar T_R\,,
\end{equation}
while we parametrize the transformation of $T_L$ in a general way as\footnote{To be precise, this transformation only spans the $\SO{2}\simeq\U{1}$ subgroup of the most general $\mathrm{O}(2)$ transformation, however our conclusions also hold in the general case.}
\begin{equation}
    \bar T_L\to \sqrt{i}\,e^{-i\varphi}\,\bar T_L\,,\qquad \, T_L\to \sqrt{i}\,e^{i\varphi}\, T_L\,,
\end{equation}
for some phase $\varphi$, in agreement with \eqref{eq:trafo fermions}.
We find that vector-like mass terms are not allowed for any transformation of $T_L$ and therefore goofy symmetry generically implies $m_{\cal T}=m=0$ (which remains in fact true for the most generic goofy transformation of $T_R$).

The vector-like fermions can get masses from spontaneous breaking of goofy invariance by coupling them for example to the scalar singlet $S$ from Sec.~\ref{sec:singl}. If $T_L$ transforms as 
\begin{equation}
    \bar T_L\to -i\sqrt{i}\,\bar T_L\,,\qquad \, T_L\to i\sqrt{i}\, T_L\,,
\end{equation}
then the new Yukawa interactions
\begin{equation}
    {\cal L} \supset y_S S \bar T_L T_R + y_S^\ast S^\ast \bar T_R T_L\,.
\end{equation}
as well as similar interactions with $T_R\to u_R$ are goofy invariant. The vev of $S$ generates a mass for the vector-like quark of\footnote{For simplicity, we omit couplings involving both the new quark and the right-handed up quark.} 
\begin{equation}
    m_T \simeq \frac{|y_S|}{\sqrt 2}v_S\,,
\end{equation}
guaranteeing that the masses of the new fermions are limited by the scale of spontaneous goofy symmetry breaking. Since naturalness of the Higgs mass suggests spontaneous goofy breaking around the $\mathcal{O}(\mathrm{TeV})$ scale, fermionic loop corrections can then not introduce problematically large corrections to the Higgs mass.

\subsection{Weinberg operator}
Majorana neutrino masses are commonly generated via the $D=5$ Weinberg operator. This operator transforms as
\begin{equation}
\begin{split}
    \mathcal{L}=&\frac{c}{\Lambda}(\bar {L_L}\tilde H)(\tilde H^T L_L^c)+\frac{c^*}{\Lambda}(\bar{L_L^c}\tilde H^*)(\tilde H^\dagger L_L)\\
    &\to-i\frac{c}{\Lambda}(\bar {L_L}\tilde H)(\tilde H^T L_L^c)-i\frac{c^*}{\Lambda}(\bar{L_L^c}\tilde H^*)(\tilde H^\dagger L_L)\,,
\end{split}
\end{equation}
where we used the same transformation for the lepton doublets as previously for quark doublet (see \eqref{fermgoof}) which implies $L^c_L=-i(\bar{L_L}\gamma^0\gamma^2)^T\to \sqrt{i}\, L_L^c$ and $\bar{L_L^c}=iL_L^T\gamma^0\gamma^2\to\sqrt{i}\,\bar{L_L^c}$.
Therefore the Weinberg operator violates goofy symmetry and thus vanishes unless goofy symmetry is spontaneously broken in which case $\Lambda$ is related to the scale of spontaneous goofy symmetry breaking. In the latter case, the generation of neutrino masses would be intricately linked to the generation of the Higgs mass term.

All this suggests that goofy symmetries might play an important role in model building by constraining the set of possible theories.

\subsection{Spontaneous goofy symmetry breaking via dimensional transmutation}

Goofy symmetry might be broken via dimensional transmutation, in which case the breaking scale is exponentially suppressed compared to high scales, such as the Planck scale $M_\mathrm{Pl}$. Such a scenario could be realized via a condensate of a new strong interaction (see, e.g., \cite{Hill:2002ap}).\footnote{Since goofy symmetry tends to forbid scalar mass terms, an alternative~\cite{Trautner:2025yxz} could be a Coleman Weinberg mechanism~\cite{Coleman:1973jx} in an extended massless scalar sector.} In case of a condensate of a scalar bilinear \mbox{${\cal O}=\phi^\dagger \phi$}, which is not invariant under goofy transformations, $\langle {\cal O} \rangle\!=\!\Lambda_\phi^2\!>\! 0$, spontaneously breaks goofy symmetry. The goofy invariant portal term to the Higgs doublet
\begin{equation}
{\cal L} \supset
- \lambda_p|\phi|^2|H|^2\,,
\end{equation}
could thereby introduce a Higgs mass of $m_H^2=\lambda_p \Lambda_\phi^2$. In case of a fermion condensate $\langle\bar\psi\psi\rangle=\Lambda_\psi^3$, the Higgs mass can be generated via the goofy invariant $D=8$ operator
\begin{equation}
    \mathcal{L}\supset \frac{c}{\Lambda^4}\bar\psi\psi\,\bar\psi\psi\, |H|^2\,.
\end{equation}
Such a construction could explain the smallness of spontaneous goofy-symmetry breaking via dimensional transmutation, leading to $\Lambda_{\phi,\psi} \ll M_{\rm Pl}$, while the absence of explicit mass terms in the first place would be guaranteed by goofy symmetry.

\section{Conclusion}

In this letter, we have applied goofy transformations to the SM, thereby also extending them to include the fermion sector.
We have shown that all SM interactions are invariant under the appropriate goofy transformations, while the Higgs-mass term is forbidden. Goofy symmetry can thus provide a new means to solve the hierarchy problem.
Since, unlike shift symmetry, it is not broken explicitly by SM interactions, goofy symmetry renders a small Higgs mass technically natural in the 't Hooft sense. 

We also explored simple goofy-symmetric extensions of the SM, including a scalar singlet, vector-like fermions, and higher-dimensional operators, demonstrating how massive states can emerge via spontaneous goofy symmetry breaking. A typical prediction would be new states related to this symmetry breaking, such as a scalar singlet around the TeV scale. This could also furnish a dark matter candidate. 

It would be interesting to explore how further puzzles could be addressed in (extensions of) the goofy-symmetric SM. Examples are the generation of a viable baryon asymmetry (potentially via an enlarged scalar sector, such as in \ref{sec:singl}, enhancing the electroweak phase transition), the strong CP problem, the generation of small neutrino masses, and the question of grand unification.
Finally, also the mathematical foundations of goofy transformations, including connections to spacetime symmetries, and 
the mechanism of spontaneous goofy symmetry breaking should be further scrutinized on the way of establishing them as a route in model building.

\section*{Acknowledgments}
We are grateful to Tim Herbermann, Jisuke Kubo, Manfred Lindner, Markus Reinig and Benhao Tang for useful discussions and to Yi Chung and Andreas Trautner for insightful comments on the manuscript.

\twocolumngrid
\bibliographystyle{utphys}
\bibliography{bib.bib}

\providecommand{\href}[2]{#2}\begingroup\raggedright\begin{thebibliography}{10}

\bibitem{Ferreira:2023dke}
P.~M. Ferreira, B.~Grzadkowski, O.~M. Ogreid, and P.~Osland, ``{New symmetries of the two-Higgs-doublet model},'' \href{http://dx.doi.org/10.1140/epjc/s10052-024-12561-8}{{\em Eur. Phys. J. C} {\bfseries 84} no.~3, (2024) 234}, \href{http://arxiv.org/abs/2306.02410}{{\ttfamily arXiv:2306.02410 [hep-ph]}}.

\bibitem{FerreiraTalk}
P.~M. Ferreira, ``{New Symmetries of the Two Higgs Doublet Model}.'' {Presentation at {\it Higgs as a Probe of New Physics 2023}}.
\newblock Osaka, Japan, 5th June, 2023.

\bibitem{Trautner:2025yxz}
A.~Trautner, ``{Goofy is the new Normal},'' \href{http://arxiv.org/abs/2505.00099}{{\ttfamily arXiv:2505.00099 [hep-ph]}}.

\bibitem{Haber:2025cbb}
H.~E. Haber and P.~M. Ferreira, ``{RG-stable parameter relations of a scalar field theory in absence of a symmetry},'' \href{http://dx.doi.org/10.1140/epjc/s10052-025-14148-3}{{\em Eur. Phys. J. C} {\bfseries 85} no.~5, (2025) 541}, \href{http://arxiv.org/abs/2502.11011}{{\ttfamily arXiv:2502.11011 [hep-ph]}}.

\bibitem{Ferreira:2025ate}
P.~M. Ferreira, B.~Grzadkowski, and O.~M. Ogreid, ``{Imaginary scaling},'' \href{http://arxiv.org/abs/2506.21145}{{\ttfamily arXiv:2506.21145 [hep-ph]}}.

\bibitem{Pilaftsis:2024uub}
A.~Pilaftsis, ``{Dirac algebra formalism for Two Higgs Doublet Models: The one-loop effective potential},'' \href{http://dx.doi.org/10.1016/j.physletb.2024.139147}{{\em Phys. Lett. B} {\bfseries 860} (2025) 139147}, \href{http://arxiv.org/abs/2408.04511}{{\ttfamily arXiv:2408.04511 [hep-ph]}}.

\bibitem{Schmaltz:2005ky}
M.~Schmaltz and D.~Tucker-Smith, ``{Little Higgs review},'' \href{http://dx.doi.org/10.1146/annurev.nucl.55.090704.151502}{{\em Ann. Rev. Nucl. Part. Sci.} {\bfseries 55} (2005) 229--270}, \href{http://arxiv.org/abs/hep-ph/0502182}{{\ttfamily arXiv:hep-ph/0502182}}.

\bibitem{Bally:2022naz}
A.~Bally, Y.~Chung, and F.~Goertz, ``{Hierarchy problem and the top Yukawa coupling: An alternative to top partner solutions},'' \href{http://dx.doi.org/10.1103/PhysRevD.108.055008}{{\em Phys. Rev. D} {\bfseries 108} no.~5, (2023) 055008}, \href{http://arxiv.org/abs/2211.17254}{{\ttfamily arXiv:2211.17254 [hep-ph]}}.

\bibitem{Chung:2024zsr}
Y.~Chung, A.~Bally, and F.~Goertz, ``{Looking for the solution to the Hierarchy Problem in Top physics},'' \href{http://dx.doi.org/10.22323/1.476.0343}{{\em PoS} {\bfseries ICHEP2024} (2025) 343}.

\bibitem{Chung:2023gcm}
Y.~Chung and F.~Goertz, ``{Third-generation-philic hidden naturalness},'' \href{http://dx.doi.org/10.1103/PhysRevD.110.115019}{{\em Phys. Rev. D} {\bfseries 110} no.~11, (2024) 115019}, \href{http://arxiv.org/abs/2311.17169}{{\ttfamily arXiv:2311.17169 [hep-ph]}}.

\bibitem{Goertz:2015dba}
F.~Goertz, ``{Electroweak Symmetry Breaking without the $\mu^2$ Term},'' \href{http://dx.doi.org/10.1103/PhysRevD.94.015013}{{\em Phys. Rev. D} {\bfseries 94} no.~1, (2016) 015013}, \href{http://arxiv.org/abs/1504.00355}{{\ttfamily arXiv:1504.00355 [hep-ph]}}.

\bibitem{ATLAS:2022jtk}
{\bfseries ATLAS} Collaboration, G.~Aad {\em et~al.}, ``{Constraints on the Higgs boson self-coupling from single- and double-Higgs production with the ATLAS detector using pp collisions at s=13 TeV},'' \href{http://dx.doi.org/10.1016/j.physletb.2023.137745}{{\em Phys. Lett. B} {\bfseries 843} (2023) 137745}, \href{http://arxiv.org/abs/2211.01216}{{\ttfamily arXiv:2211.01216 [hep-ex]}}.

\bibitem{ATLAS:2024xcs}
{\bfseries ATLAS} Collaboration, G.~Aad {\em et~al.}, ``{Search for triple Higgs boson production in the 6b final state using pp collisions at s=13\,\,TeV with the ATLAS detector},'' \href{http://dx.doi.org/10.1103/PhysRevD.111.032006}{{\em Phys. Rev. D} {\bfseries 111} no.~3, (2025) 032006}, \href{http://arxiv.org/abs/2411.02040}{{\ttfamily arXiv:2411.02040 [hep-ex]}}.

\bibitem{DiMicco:2019ngk}
J.~Alison {\em et~al.}, ``{Higgs boson potential at colliders: Status and perspectives},'' \href{http://dx.doi.org/10.1016/j.revip.2020.100045}{{\em Rev. Phys.} {\bfseries 5} (2020) 100045}, \href{http://arxiv.org/abs/1910.00012}{{\ttfamily arXiv:1910.00012 [hep-ph]}}.

\bibitem{Helset:2022pde}
A.~Helset, E.~E. Jenkins, and A.~V. Manohar, ``{Renormalization of the Standard Model Effective Field Theory from geometry},'' \href{http://dx.doi.org/10.1007/JHEP02(2023)063}{{\em JHEP} {\bfseries 02} (2023) 063}, \href{http://arxiv.org/abs/2212.03253}{{\ttfamily arXiv:2212.03253 [hep-ph]}}.

\bibitem{Sartore:2020gou}
L.~Sartore and I.~Schienbein, ``{PyR@TE 3},'' \href{http://dx.doi.org/10.1016/j.cpc.2020.107819}{{\em Comput. Phys. Commun.} {\bfseries 261} (2021) 107819}, \href{http://arxiv.org/abs/2007.12700}{{\ttfamily arXiv:2007.12700 [hep-ph]}}.

\bibitem{Hill:2002ap}
C.~T. Hill and E.~H. Simmons, ``{Strong Dynamics and Electroweak Symmetry Breaking},'' \href{http://dx.doi.org/10.1016/S0370-1573(03)00140-6}{{\em Phys. Rept.} {\bfseries 381} (2003) 235--402}, \href{http://arxiv.org/abs/hep-ph/0203079}{{\ttfamily arXiv:hep-ph/0203079}}. [Erratum: Phys.Rept. 390, 553--554 (2004)].

\bibitem{Coleman:1973jx}
S.~R. Coleman and E.~J. Weinberg, ``{Radiative Corrections as the Origin of Spontaneous Symmetry Breaking},'' \href{http://dx.doi.org/10.1103/PhysRevD.7.1888}{{\em Phys. Rev. D} {\bfseries 7} (1973) 1888--1910}.

\end{thebibliography}\endgroup

\end{document}